\documentclass[prl,showpacs,twocolumn]{revtex4}
\usepackage{graphicx}

\begin{document}
\bibliographystyle{prsty}

\title{Miniature plasmonic wave plates}
\author{Aur\'{e}lien Drezet}
\author{Cyriaque Genet}
\author{Thomas W.~Ebbesen}

\affiliation{Isis, Louis Pasteur University, 8 all\'{e}e Gaspard
Monge, 67000, Strasbourg, France}

\date{\today}

\begin{abstract}
Linear birefringence, as implemented in wave plates, is a natural
way to control the state of polarization of light. We report on a
general method for designing miniature planar wave plates using
surface plasmons. The resonant optical device considered here is a
single circular aperture surrounded by an elliptical antenna
grating. The difference in short and long axis of each ellipses
introduces a phase shift on the surface plasmons which enables the
realization of a quarter wave plate. Furthermore, the experimental
results and the theoretical analysis show that the general
procedure used does not influence the optical coherence of the
polarization state and allows us to explore completely the surface
of the unit Poincar\'{e} sphere by changing only the shape of the
elliptical grating.
\end{abstract}
 \maketitle

\noindent Surface plasmon polaritons (SPPs), electromagnetic
surface waves existing at the interface between a dielectric and a
metal~\cite{Raether-Book}, are particularly sensitive to tiny
variations in their local electronic environments. This creates
new opportunities and applications for photonics~\cite{Genet-2007}
by simply texturing a metal surface. For example, metal films
structured with two dimensional subwavelength hole arrays present
remarkable properties such as the extraordinary optical
transmission (EOT) which is a clear signature of SPP-light
interaction~\cite{Ebbesen-1998,Barnes-2004,Moreno-2001}.
In this particular context, several studies have started to address polarization issues, discussing in this respect the influence of the individual hole shapes. Elliptical or rectangular apertures can behave like polarizers, following the Malus
law of absorption~\cite{Gordon-2004,Zayats-2004,Degiron-2004} (see also ref.~\cite{Franz}
for similar work on elliptical nanoparticles). However these structures do not display linear birefringence.
Linear birefringence is absolutely central in optics since it allows full control of the state of polarization (SOP) of
light without absorption. A half wave-plate rotates the plane of polarization while a quarter wave-plate converts linear
polarized light into a circular one, the combination of the two enabling a complete exploration of all polarization states.\\
\indent In this letter, we report for the first time, both
experimentally and theoretically, the design and characterization
of a plasmonic optical wave-plate. In order to obtain the linear
birefringence, we have developped a modified version of the single
circular nano-aperture surrounded by periodic circular
corrugations, also known as a \emph{bull's eye}
structure~\cite{Lezec-2002}. Such an optical grating acts as a
miniature antenna presenting huge EOT for optical wavelength
inside a narrow band centered on the SPP
resonance~\cite{Lezec-2002,Laux-2007}. The specificity of the
structure presented here is its unique ability to control the SOP
of the electromagnetioc field going through the aperture. This is
achieved by introducing a well defined excentricity in the grating
geometry which in turn modifies the phase of the excited SPP and
consequently the polarization of the transmitted light. This
resembles in a wide sense the phase matching in distributed
feedback lasers (DFB). To fully characterize the optical behaviour
of our device, a genuine polarization tomography of the isolated
subwavelength aperture had to be
implemented. Furthermore we have developped a microscopic (dipolar) model to link structural design with change of SOP.  \\
\begin{figure}[hbtp]
\includegraphics[width=8cm]{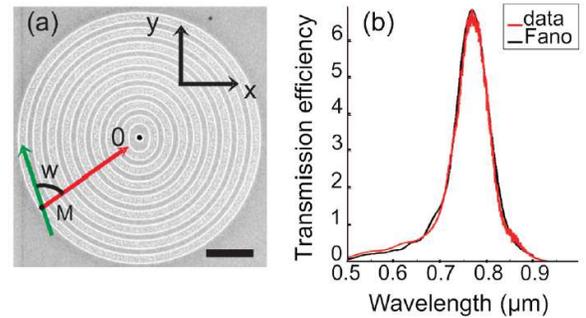}
\caption{(a) Scanning electron microscopy image of the  elliptical
bull's eye structure. The scale bar is $2\mu m$ long. The green
arrow is an excited point dipole $\mathbf{P}_M$ located in
$\mathbf{M}$. SPP (red arrow) launched in $\mathbf{M}$ propagates
along the radial direction $\mathbf{MO}$ to reach the hole located
in $\mathbf{O}$. $w$ is the angle between $\mathbf{MO}$ and
$\mathbf{P}_M$. (b) Comparison between the experimental white
light transmission spectrum (red curve) associated with the
structure shown in (a) and the theoretical prediction (black
curve) obtained with the 2D dipoles model. The transmission scale
is dimensionless and corresponds to the ratio between the
transmitted and the incoming power at the level of the aperture.
}
\end{figure}
Beside its fundamental interest, such control over the SOP can be
used broadly in photonic applications requiring local addressing,
e.~g., detectors~\cite{Ishi-2005}, displays~\cite{Laux-2007} and
compact circular polarization antennas~\cite{antenna}. In
addition, ultrafast opto-magnetic data storage has been
demonstrated with femtosecond lasers and circularly polarized
light~\cite{Stanciu-2007}. The device demonstrated here creates
both the right helicity and the large fields in a
tiny volume favorable for such purposes.\\
\indent For our experiments we consider a bull's eye structure
made of 8 grooves and fabricated by FIB milling in a 300 nm thick
Au film (Fig.~1). The hole diameter is 260 nm and the grooves
width and depth are 370 and 80 nm respectively. The groove shape
is chosen to be elliptical with the long axis $a_{n}=n\cdot P+P/4$
and the short axis $b_{n}=n\cdot P$. Here $P= 760$ nm is the
period of the grating (which equals the SPP wavelength
$\lambda_{SPP}$ for a laser excitation at 785 nm~\cite{JC-1972})
and $n$ is an integer going from 1 to 8 (see Fig.~1(a)). Also
shown in Fig.~1(b) is the transmission spectra of the structure
with a resonant peak at $\lambda_0\simeq 777 nm$. The measured
extraordinary transmission efficiency (larger than one) is a
direct signature of the involvement of SPP~\cite{Genet-2007}. The
presence of this transmission peak proves that despite the small
increment of $\delta L=P/4$ between the long and short axis of the
ellipses the structure still behaves like a miniature
antenna.\begin{figure}[hbtp]
\includegraphics[width=8cm]{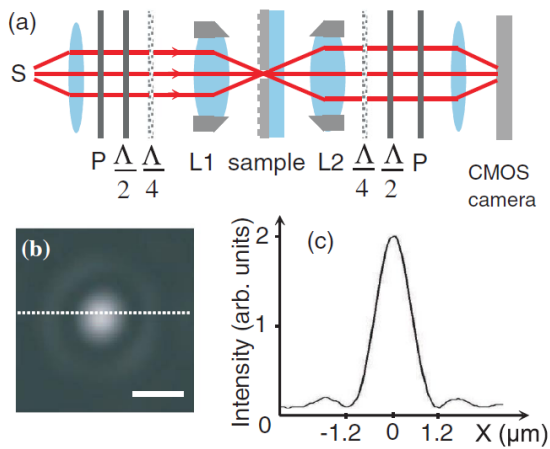}
\caption{(a) Sketch of the  optical polarization tomography setup.
$P$, $\Lambda/4$, $\Lambda/2$ are  respectively polarizers,
quarter wave plates, and half wave plates located in the input and
output beams. The image is recorded with a CMOS camera and the
light source $S$ is a laser diode emitting at $\lambda= 785 nm$.
$L_1$ and $L_2$ are two objective lenses  ($\times 50$, $NA=0.55$)
and ($\times 40$, $NA=0.6$) respectively. (b) Typical camera image
of the light transmitted through our structure (the scale bar is 2
$\mu m $). (c) Cross cut of the intensity profile along the white
dotted line shown in (b).}
\end{figure} We can justify our choice for the grating symmetry
on theoretical grounds. In our model we discretize the grooves
into a sum of point dipoles $\mathbf{P}_M$ proportional to the
local electric field at M. Each dipole is excited coherently by
the light impinging normal to the metal film and SPPs are launched
in the direction of the central nanohole where they excite an
in-plane radiating dipole~\cite{Genet-2004}. To reproduce
completely the system, we introduce a second transmission channel
in which the central dipole is excited directly by the incident
light. The interference between these two channels leads to a Fano
like effect~\cite{Genet-2003} resulting in the observed
transmission peak. The relative (complex) amplitude between these
two channels was fitted to reproduce the spectra of
Fig.~1(b)~\cite{note-2}. The good agreement between our model and
the data (see Fig.~1(b)) allows us to use it for predicting the
optical behavior of the structure at a given $\lambda$. The
principle of the device can be illustrated by considering only the
point dipoles located along the short and long axes of the
ellipses. It is thus clear that $\delta L$ corresponds to a phase
shift $\phi_{SPP}=2\pi \delta L/\lambda_{SPP}=\pi/2$ between SPPs
propagating along the long (y) and the short (x) axes.
Additionally the coupling between the incident light and SPPs
depends on the cosine of the angle $w$ between the radial vector
$\textbf{MO}$ and $\mathbf{P}_M$ (Fig.~1 (a)). It means that if
the incident linear polarization is switched from a direction
parallel to the x axis to a direction parallel to the y axis then
the radiating central dipole will change from $\alpha
\hat{\mathbf{x}}$ (where $\alpha$ is a constant) to
$e^{i\pi/2}\alpha \hat{\mathbf{y}}$. From the point of view of
this idealized picture (which neglects damping) we deduce that the
system behaves like a birefringent biaxial medium, i.~e., a
perfect quarter wave plate, with fast and slow axes parallel
respectively to the x and y axes. Obviously if we now take into
account all the dipoles as well as the Fano interference effect
and the finite value of the SPP propagation length $L_{SPP}$
(damping) in the structure the actual
result will naturally deviate from this idealized case~\cite{Kim-2003,note-L}.\\
\indent In order to study experimentally the SOP conversion by our
structure, we carried out a complete polarization
tomography~\cite{Jeune-97} using the optical setup sketched in
Fig.~2(a). A laser beam at $\lambda=785 nm$ is focused normally
onto the structure by using an objective $L_1$. The transmitted
light is collected by a second objective $L_2$ forming an Airy
spot on the camera (see Fig.~(2b-c)) as expected since the hole
behaves like a point source in a opaque gold film. In our
experiments, the intensity is thus defined by taking the maximum
of the Airy spot shown on Fig.~2(b). The SOP of light is prepared
and analyzed with half wave plates, quarter wave plates, and
polarizers located before and after the
objectives~\cite{Jeune-97,Born,Yuri}. We recall that the complete
knowledge of the SOP require 6 intensity projection measurements
$\langle I_{\hat{\mathbf{a_i}}}\rangle= \langle
|\mathbf{E}\cdot\hat{\mathbf{a_i}}|^2\rangle$ made along the 4
linear polarization vectors $\hat{\mathbf{x}}$,
$\hat{\mathbf{y}}$,
$\hat{\mathbf{p}}=(\hat{\mathbf{x}}+\hat{\mathbf{y}})/\sqrt{2}$,
$\hat{\mathbf{m}}=(\hat{\mathbf{x}}-\hat{\mathbf{y}})/\sqrt{2}$,
and along the two circular polarization vectors
$\hat{\mathbf{L}}=(\hat{\mathbf{x}}+i\hat{\mathbf{y}})/\sqrt{2})$,
$\hat{\mathbf{R}}=(\hat{\mathbf{x}}-i\hat{\mathbf{y}})/\sqrt{2})$.
It is convenient~\cite{Born} to introduce the four Stokes
parameters $S_1=\langle I_{\hat{\mathbf{x}}}-I_{\hat{\mathbf{y}}}
\rangle$ $S_2=\langle I_{\hat{\mathbf{p}}}-I_{\hat{\mathbf{m}}}
\rangle$, $S_3=\langle I_{\hat{\mathbf{L}}}-I_{\hat{\mathbf{R}}}
\rangle$, and $S_0=\langle
I_{\hat{\mathbf{x}}}+I_{\hat{\mathbf{y}}}\rangle=\langle
I_{\textrm{total}}\rangle$. The goal of this polarization
tomography is then the determination of the $4\times 4$ Mueller
matrix $\mathcal{M}$ characterizing the transformation of the
input Stokes parameters during the interaction of the laser light
with the structure. In order to write down the full Mueller
matrix, we measured $6\times 6$ intensity projections
corresponding to the 6 previously mentioned
unit vectors for the input and the output polarizations~\cite{note-Mueller}.\\
\indent At first, the isotropy of the bare setup was checked by
measuring the Mueller matrix $\mathcal{M}^{\textrm{glass}}$ with a
glass substrate. Up to a normalization constant, we deduced that
$\mathcal{M}^{\textrm{glass}}$ is practically identical to the
identity matrix $\mathcal{I}$ with individuals elements deviating
from it by no more than 0.02. It implies that the optical setup
does not induce depolarization and that consequently we can rely
on our measurement procedure for obtaining $\mathcal{M}$. Optical
depolarization (i.~e, losses in polarization coherence) can be
precisely quantified through the degree of purity of the Mueller
matrix defined by \cite{Gill-2000}
$D\left(\mathcal{M}\right)=\left(\frac{\textrm{Tr}[\mathcal{M}^{\dagger}\mathcal{M}]-\mathcal{M}_{00}^{2}}{3\mathcal{M}_{00}^{2}}
\right)^{1/2}\leq 1. $ We find
$D\left(\mathcal{M}^{\textrm{glass}}\right)=0.9851$. We impute the
residual depolarization ($1-D\sim$2\%) to the lenses and to
alignment errors. It should be noted that the incident
illumination spot size on the sample was varied between 2 and 20
$\mu$m without affecting the matrix, i.~e., without introducing
additional depolarization. In the following experiment, we
consider the case of a large gaussian spot with FWHM=20 $\mu$m in
order to illuminate the whole structure~\cite{note-spot}. We then
measured the Mueller matrix of our structure and found that:
\begin{eqnarray}
\mathcal{M}^{\textrm{exp.}}= \left(\begin{array}{cccc}
    \underline{1.000} & \underline{0.107} &  -0.008   &0.000   \\
    \underline{0.111} &   \underline{0.972} &-0.002    & -0.004  \\
    0.004 & -0.002  &  \underline{0.306}  & \underline{-0.932}\\
  0.001  & -0.017  &  \underline{0.934}  &  \underline{0.294}
\end{array}\right)
\end{eqnarray}
which is clearly block diagonal, up to experimental errors. It is
also remarkable that we have
$D\left(\mathcal{M}^{\textrm{exp.}}\right)=0.981$. This means that
despite the existence of the SPP transmission channel, the
polarization coherence is not lost during the propagation through
the structure. This situation contrasts with previous SOP
tomography measurements on metallic hole arrays in which the
polarization degrees of freedom were mixed with spatial
information responsible for SPP-induced
depolarization~\cite{Altewischer-2005}. Beside these two points,
the matrix $\mathcal{M}^{\textrm{exp.}}$ exhibits several
interesting symmetrical features which relate to the polarization
properties of the device. First, it can be observed that in our
experimental procedure the polarization in the Airy spot (see
Fig.~2(b)) is homogeneous~\cite{note-3}. This means that in our
analysis we are actually doing the polarization tomography of the
central radiating dipole, i.~e., we are dealing only with the
SU(2) point symmetry of the Mueller matrix. In such context, the
rectangular point symmetry group $C_{2v}$ of the ellipse imposes
that the ($2\times 2$) Jones matrix~\cite{Born} $\mathcal{J}$
connecting the incident electric field
$\left(E_{x}^{\textrm{in}},E_{y}^{\textrm{in}}\right)$ to the
transmitted electric field
$\left(E_{x}^{\textrm{out}},E_{y}^{\textrm{out}}\right)$ must be
diagonal in the x and y basis, i.~e.~,
$\mathcal{J}\propto\left(\begin{array}{cc} 1 & 0\\0 &\beta
\end{array}\right)$ where $\beta=\rho e^{i\phi}$ is a complex
number. In analogy with bulk optics, $\rho$ and $\phi$ measure
respectively the relative dichroism (i.~e.~the relative
absorbtion) and the birefringence of this biaxial 2D medium.
Clearly $\phi$ is reminiscent of $\phi_{SPP}$ discussed above.
Using $\mathcal{J}$ we obtain the theoretical Mueller matrix
\begin{eqnarray}
\mathcal{M}^{C_{2v}}\propto\left(\begin{array}{cccc}
    1+\rho^2 & 1-\rho^2&  0  &  0\\
     1-\rho^2&  1+\rho^2&  0  & 0 \\
    0 & 0  &  2\textrm{Re}\left(\beta\right) & -2\textrm{Im}\left(\beta\right)\\
    0 &  0 &  2\textrm{Im}\left(\beta\right)  &  2\textrm{Re}\left(\beta\right)
\end{array}\right)
\end{eqnarray}
which is  similar to $\mathcal{M}^{\textrm{exp.}}$ and in
particular satisfies the symmetries
$\mathcal{M}_{01}=\mathcal{M}_{10}$,
$\mathcal{M}_{00}=\mathcal{M}_{11}$,
$\mathcal{M}_{22}=\mathcal{M}_{33}$, and
$\mathcal{M}_{23}=-\mathcal{M}_{32}$ observed experimentally. We
deduce
$\rho=\left(\frac{\mathcal{M}_{00}-\mathcal{M}_{01}}{\mathcal{M}_{00}+\mathcal{M}_{01}}\right)^{1/2}$
and $\tan{\phi}=\mathcal{M}_{32}/\mathcal{M}_{22}$. Using Eqs.~1,2
we obtain $\rho\simeq 0.898$ and $\phi\simeq 72.5^{\circ}$.
Reciprocally by injecting the previous values for $\rho$ and
$\phi$ in $\mathcal{M}^{C_{2v}}$  the result do not differ from
$\mathcal{M}^{\textrm{exp.}}$ by more than $2\%$, in agreement
with the value obtain for the residual depolarization. Then, using
the fitting parameters already considered in the transmission
spectrum Fig.~1(b), we numerically calculate the Mueller matrix
predicted by the 2D dipole model and obtain
\begin{eqnarray}
\mathcal{M}^{2D}=\left(\begin{array}{cccc}
    1.000& 0.089&  0.000  &  0.000\\
    0.089&  1.000&  0.000  & 0.000 \\
    0.000 & 0.000  & 0.446& -0.890\\
    0.0000 & 0.000 &  0.890 & 0.446
\end{array}\right)
\end{eqnarray}
which is close to $\mathcal{M}^{C_{2v}}$ and
$\mathcal{M}^{\textrm{exp.}}$ and corresponds to
$D\left(\mathcal{M}^{2D}\right)=1$. This numerical model is
sensitive to small variations of the fitting parameters and the
agreement with the experiment could be probably improved by going
beyond the
paraxial approximation for the incident light~\cite{note-spot}.\\
\begin{figure}[hbtp]
\includegraphics[width=9cm]{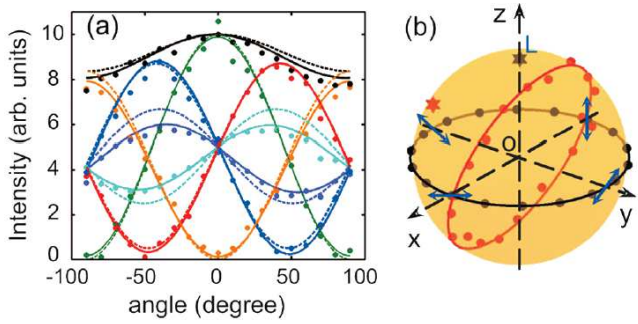}
\caption{(a) SOP analysis of the output beam for a linearly
polarized input beam. The polarization angle is measured
relatively to the x axis. Data points are compared to Eq.~1
(continuous curves) and to predictions of Eq.~3 (dotted curves).
The colors black, green, orange, cyan, magenta, red and blue
correspond respectively to $I_{\textrm{total}}$,
$I_{\hat{\mathbf{x}}}$, $I_{\hat{\mathbf{y}}}$,
$I_{\hat{\mathbf{p}}}$, $I_{\hat{\mathbf{m}}}$,
$I_{\hat{\mathbf{L}}}$, $I_{\hat{\mathbf{R}}}$. (b) Image of the
input Poincar\'{e} sphere through the transformation
$\mathcal{M}^{\textrm{exp.}}$ (yellow sphere). The black and red
data points are respectively the input and output Stokes vectors
associated with the experiment shown on (a). The circle represents
the predictions deduced from $\mathcal{M}^{\textrm{exp.}}$. The
device also converts a input $L$ state (black) star into a state
shown by a red star.}
\end{figure}
\indent Finally, we considered more closely the consequence of the
transformation defined by $\mathcal{M}^{\textrm{exp.}}$ by varying
the linear polarization $\theta$ of the input state every
$10^\circ$ from -$90^{\circ}$ to +$90^{\circ}$. On Fig.~3(a) we
show the transmitted intensity analyzed along the 6 fundamental
polarizations $\hat{\mathbf{x}}$, $\hat{\mathbf{y}}$,
$\hat{\mathbf{p}}$, $\hat{\mathbf{m}}$, $\hat{\mathbf{L}}$ and
$\hat{\mathbf{R}}$ as a function of $\theta$. The interference
fringes observed are compared with the predictions given by the 2D
dipole model (dotted curves) and with the intensity deduced from
the Mueller matrix $\mathcal{M}^{\textrm{exp.}}$ (continuous
curves). In both cases the agreement is very good showing once
again the consistency of the different measurements and
deductions. Furthermore, this can be geometrically illustrated by
using the Stokes vector~\cite{Born} defined by
$\mathbf{S}=[S_1\hat{\mathbf{x}}+S_2\hat{\mathbf{y}}+S_3\hat{\mathbf{z}}]/S_0$.
The surface drawn by the input Stokes vector is called a
Poincar\'{e} sphere and has the radius $D=1$. As shown in
Fig.~3(b) the operator $\mathcal{M}^{\textrm{exp.}}$ defines a
geometrical transformation connecting this Poincar\'{e} sphere to
an output surface with a characteristic radius
$D\left(\mathcal{M}^{\textrm{exp.}}\right)$. This experimental
surface is very close to the ideal sphere $D=1$ in agreement with
the absence of net depolarization as discussed earlier. The
experiment shown on Fig.~3(a) is also represented on this sphere.
From Eq.~2, we deduce that if the input Stokes vector explores the
equator corresponding to linear polarizations, then the output
Stokes vector draws a circle of radius $D\simeq 1$ which is
contained in the plane $y/z=\mathcal{M}_{33}/\mathcal{M}_{32}$
making the angle $90^{\circ}-\phi=17.5^{\circ}$ with the z axis.
These predictions are directly consistent with the observations.
Additionally, for a pure left circular input SOP, we
experimentally obtain
$\mathbf{S}^{\textrm{exp.}}=0.1123\hat{\mathbf{x}}-0.8984\hat{\mathbf{y}}+0.3104\hat{\mathbf{z}}$
in agreement with the value deduced from $\mathcal{M}^{C_{2v}}$:
$\mathbf{S}^{C_{2v}}=0.1067\hat{\mathbf{x}}-0.9272\hat{\mathbf{y}}+0.2949\hat{\mathbf{z}}$.\\
\indent To conclude, all this experimental and theoretical
analysis demonstrate that we have a clear understanding of the SPP
structure considered here. First, we have $\rho\simeq 1$ which
implies that the system acts essentially as a birefringent medium
with Jones Matrix $\mathcal{J}\simeq\left(\begin{array}{cc} 1 &
0\\0 &e^{i\phi}
\end{array}\right)$, i.~e., a wave plate.
Second, the value obtained for $\phi$ shows that the system
differs slightly from an ideal quarter waveplate for which
$\phi=90^{\circ}$. From the point of view of the Poincar\'{e}
sphere, this angle measures directly the inclination of the output
circle shown on Fig.~3(b). For a perfect quarter wave plate this
circle will go through the poles, i.~e., a complete conversion
from linear  to circular polarization will become possible if the
input SOP is polarized along $\hat{\mathbf{p}}$ or
$\hat{\mathbf{m}}$. In this context, numerical calculations with
the 2D dipoles model show that by changing slightly the value for
the long axis increment $\delta L$ we can change the phase $\phi$
continuously. This means than with such SPP device we can in
principle tailor and generate any kind of SOP conversion on the
Poincar\'{e} sphere going from the equator (linear polarization)
to the poles (circular polarization) or vice versa. We expect that
the SPP control over the polarization presented in this letter could have many applications in  photonics and in  information storage technology.\\
The authors acknowledge financial support from the EC under
project No.~IST-FP6-034506.


\begin{references}
\bibitem{note-2}
Additionally we modelled the transmission spectra of the cental
isolated hole alone by using a fit of experimental spectrum.
\bibitem{note-Mueller}
Actually only 16 measurements are needed to determine
$\mathcal{M}$~ \cite{Jeune-97}. Our systematic procedure is thus
more than sufficient to obtain $\mathcal{M}$.
\bibitem{note-spot}
For spot size $FWHM\simeq 20$ $\mu$m the effective numerical
aperture of the incident beam at $\lambda=785$ nm is only
$\lambda/FWHM\simeq 0.04$. At this low numerical aperture regime,
the parameters used for the fit of the 2D dipole model at normal
incidence are still pertinent.
\bibitem{note-3}
Given the fact that $\mathcal{M}^{\textrm{glass}}\simeq
\mathcal{I}$, this is directly verified by using a pair of crossed
polarizers in the input and output beam, i.~e., measuring the
extinction.
\end{references}
\end{document}